\documentclass{IEEEtran}
\usepackage{framed,multirow}
\usepackage{cite}
\usepackage{amsmath,amssymb,amsfonts}
\usepackage{algorithmic}
\usepackage{graphicx}
\usepackage{textcomp}
\usepackage{multirow}
\usepackage{comment}
\usepackage{tabu}
\usepackage{makecell}
\usepackage[colorlinks=true]{hyperref}
\usepackage[table]{xcolor}
\usepackage{todonotes}
\usepackage{enumitem}

\def\ie{\emph{i.e.}}

\usepackage{amssymb}
\usepackage{latexsym}
\usepackage{bbding}
\usepackage{pifont}
\usepackage{wasysym}
\usepackage{url}
\usepackage{xcolor}

\usepackage{hyperref}

\definecolor{newcolor}{rgb}{.8,.349,.1}

\def\citep{\cite}
\def\citet{\cite}

\begin{document}
\title{Spatial Attention-based Implicit Neural Representation for Arbitrary Reduction of MRI Slice Spacing}
\author{Xin Wang\IEEEauthorrefmark{1}, %
	Sheng Wang\IEEEauthorrefmark{1}, %
	Honglin Xiong, %
	Kai Xuan, %
	Zixu Zhuang, %
	Mengjun Liu, %
        Zhenrong Shen, %
        Xiangyu Zhao, %
        Lichi Zhang\IEEEauthorrefmark{2}, %
	and Qian Wang\IEEEauthorrefmark{2}
\thanks{\IEEEauthorrefmark{1} Both authors contributed equally to this work.
}
\thanks{\IEEEauthorrefmark{2} Corresponding author.
}
\thanks{Xin Wang, Sheng Wang, Kai Xuan, Zixu Zhuang, Mengjun Liu, Zhenrong Shen, Xiangyu Zhao, and Lichi Zhang are with School of Biomedical Engineering,
Shanghai Jiao Tong University, Shanghai 200030, China
(e-mail: \{wangxin1007, wsheng, kaixuan, zixuzhuang, mjliu2020, zhenrongshen, xiangyu.zhao, lichizhang\}@sjtu.edu.cn).}
\thanks{Honglin Xiong and Qian Wang are with School of Biomedical Engineering, ShanghaiTech University, Shanghai 201210, China. 
(e-mail: \{xionghl, wangqian2\}@shanghaitech.edu.cn).}}

\maketitle






\begin{abstract}
Magnetic resonance (MR) images collected in 2D clinical protocols typically have large inter-slice spacing, resulting in high in-plane resolution and reduced through-plane resolution.
Super-resolution technique can enhance the through-plane resolution of MR images to facilitate downstream visualization and computer-aided diagnosis. 
However, most existing works train the super-resolution network at a fixed scaling factor, which is not friendly to clinical scenes of varying inter-slice spacing in MR scanning. 
Inspired by the recent progress in implicit neural representation, we propose a \textit{Spatial Attention-based Implicit Neural Representation (SA-INR)} network for arbitrary reduction of MR inter-slice spacing.
The SA-INR aims to represent an MR image as a continuous implicit function of 3D coordinates. 
In this way, the SA-INR can reconstruct the MR image with arbitrary inter-slice spacing by continuously sampling the coordinates in 3D space. 
In particular, a local-aware spatial attention operation is introduced to model nearby voxels and their affinity more accurately in a larger receptive field.
Meanwhile, to improve the computational efficiency, a gradient-guided gating mask is proposed for applying the local-aware spatial attention to selected areas only.
We evaluate our method on the public HCP-1200 dataset and the clinical knee MR dataset to demonstrate its superiority over other existing methods. 
\end{abstract}



\section{Introduction}

\label{sec:introduction}
Magnetic resonance (MR) imaging plays an important role in assessing and diagnosing numerous disorders due to its non-invasive and radiation-free nature, as well as its superior contrast for soft tissues.
In clinical practice, 2D scanning protocols are widely applied due to their short scanning time~\citep{mri}. However, the MR images acquired from 2D protocols typically have large inter-slice spacing compared to the fine-grained intra-slice spacing.
Therefore, it is often necessary to resample the acquired 3D volume, e.g., to make the inter-slice spacing close to the intra-slice spacing. 
In this way, the processed images not only facilitate 3D rendering, but also meet the requirements of near-isotropic resolution by many automatic image analysis tools. 


The above aim can be traditionally achieved in the way of image resampling. For example, by using trilinear interpolation, one can easily change the size and the spacing of a 3D image volume. However, the interpolation simply calculates the missing slice as a weighted average of the adjacent slices, which inevitably yields blurring and unrealistic results.

\begin{figure}[t]
    \centering
    \includegraphics[width=0.48\textwidth]{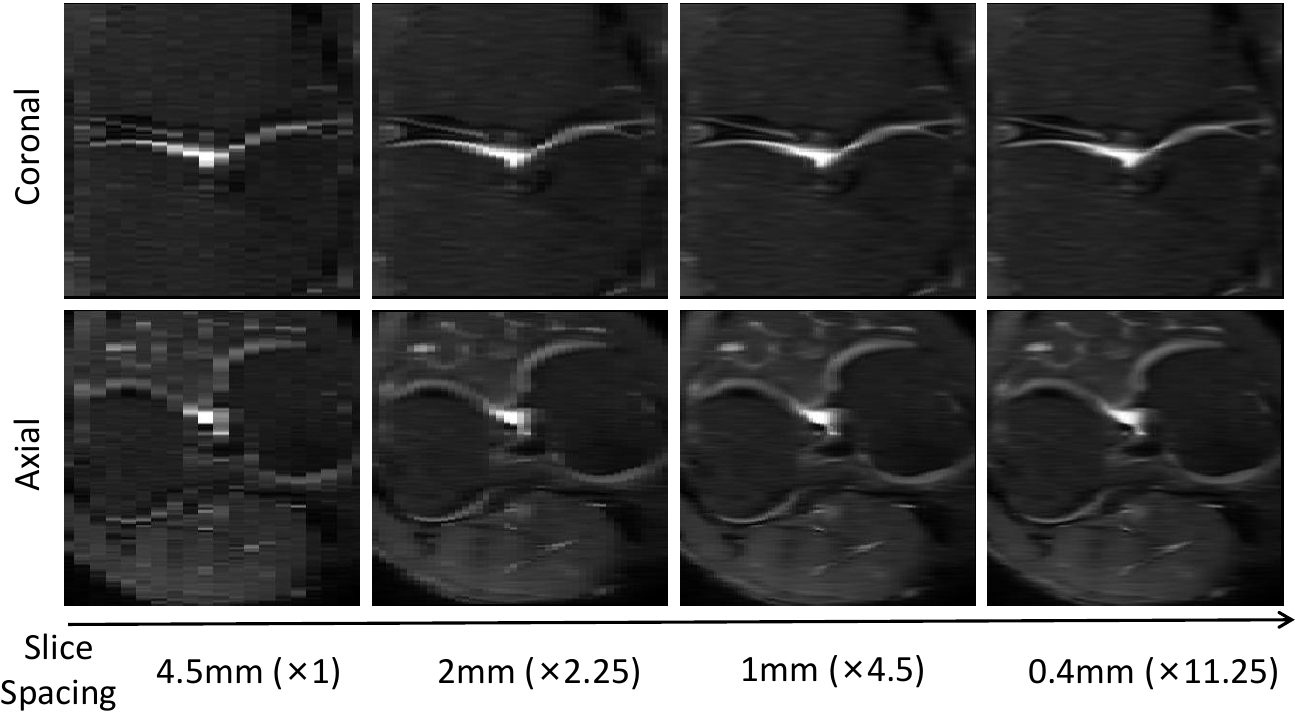}
    \caption{An instance of applying our proposed method for arbitrary reduction of knee MR slice spacing. 
    The first row and the second row show the coronal and axial views, respectively. The original slice spacing is 4.5mm (in the first column), and the voxel spacing in the sagittal plane is 0.4mm$\times$0.4mm. The through-plane resolution is enhanced progressively with the reduced slice spacing. 
    Note that our single model can succeed for various scaling factors in parentheses.
     }
    \label{fig1}
\end{figure}

Alternatively, a lot of super-resolution (SR) methods powered by machine learning have been proposed to address the issue of large inter-slice spacing for MR images~\citep{ZHANG2017531,DeepResolve,smore,kai,TSCNet}.
In this paper, we refer to the acquired MR image with large slice spacing as the low-resolution (LR) image and the image with small slice spacing as the high-resolution (HR) image for convenience.
The SR methods can either train a 3D convolutional neural network (CNN)~\citep{DeepResolve,TSCNet} or 2D~\citep{smore}, such that the missing slices in the through-plane direction can be reconstructed or interpolated from the existing slices.

Although the learning-based SR methods have demonstrated their superior capability of reducing MR slice spacing, two issues have been mostly neglected.
\begin{enumerate}
    \item Most SR networks only consider integer scaling factors, e.g., ``$\times2$'', ``$\times3$'', ``$\times4$'' to reduce the slice spacing to $1/2$, $1/3$, $1/4$ of the original. 
    However, in clinical scans, the ratio of the inter-slice spacing to the intra-slice spacing is often a non-integer. For example, in Fig.~\ref{fig1}, when the intra-slice spacing is 0.4mm$\times$0.4mm and the inter-slice spacing is 4.5mm, it may lead to the upscaling factor of $4.5/0.4 = 11.25$ to reach an isotropic voxel spacing.

    \item Most SR networks are trained at fixed scaling factors.
    For example, given a model trained with paired images of 2mm and 1mm inter-slice spacing, it is hard to generalize from the ``$\times2$'' task to another task such as ``$\times3$''. 
    On the contrary, scanning parameters in clinical can be heterogeneous in terms of inter-slice spacing. 
    Therefore, it is expected that a single network can handle such inhomogeneity in the SR tasks, instead of training a dedicated network for each specific scaling factor. Also in Fig.~\ref{fig1}, we show a test case of applying our proposed method for arbitrary reduction of knee MR slice spacing. We have successfully reconstructed the MR images of the different inter-slice spacing by using a single SR model.

\end{enumerate}

Considering the above challenges, we propose a \textit{\textbf{S}patial \textbf{A}ttention-based \textbf{I}mplicit \textbf{N}eural \textbf{R}epresentation (SA-INR)} network to interpolate slices continuously and reduce the inter-slice spacing for arbitrary factors in this paper. Our network is inspired by the recent progress in implicit neural representation for image SR~\citep{nerf,LIIF,ArSSR}.
The key idea of SA-INR is to represent an LR image as a continuous implicit function of coordinates, such that the HR image can be sampled from the continuous coordinate system and then generated from the implicit function. 

To attain the above goal, our SA-INR network has two basic modules.
First, a feature learning module is used to learn an implicit feature representation for each grid voxel in the input LR image. 
Then, a decoding module queries a new coordinate for the HR image, calculates the corresponding implicit representation based on the neighboring grid voxels in the LR image, and restores the voxel intensity from the implicit representation for the query coordinate. 
By querying all the HR coordinates for an arbitrary upsampling factor, we can generate the MR image with reduced inter-slice spacing.


It is critical to derive the implicit representation for each query coordinate in our network. 
Previous methods~\citep{LIIF,ArSSR} use linear interpolation to generate the representation of each query coordinate from its neighbors. However, the simple weighted average of four (2D) or eight (3D) neighbors may fail to capture nearby anatomic context, resulting in sub-optimal SR performance.
To better derive the corresponding implicit representation, we propose a local-aware spatial attention (LASA) operation to model the spatial dependency between each query coordinate and its neighbors. Also, LASA allows for a larger receptive field, enabling the query coordinate to include more context from a larger neighborhood.

Meanwhile, we observe that most existing SR models treat all voxels in MR images equally and perform the same calculation, regardless of their context, which is inefficient. The LASA operation should focus on the challenging areas (e.g., the areas with large intensity variation) while paying less attention to the easy regions (e.g., the flattened areas in the background).
Given that the intensity gradient can reflect the difficulty of reconstruction, we propose a gradient-guided gating mask and conditionally execute the LASA operation based on the mask. 
To enable the network to automatically learn the gating mask, the gumbel-softmax trick \citep{gumbel} is used for end-to-end training.

Furthermore, there is often no ground truth to supervise the training of the SR network in a real clinical scenario. To tackle this problem, we integrate the SA-INR network with our previous two-stage self-supervised SR framework~\citep{kai}.
First, we train a variational auto-encoder (VAE)~\citep{vae} to embed slices from real LR images. Then we synthesize HR images with reduced slice spacing by utilizing the embedding space from VAE. The synthesized HR images cannot be used as the SR output directly. Yet we can downsample them to generate LR-HR pairs with different ratios of inter-slice spacing and train the proposed SA-INR network on these pairs.
We demonstrate that our solution can effectively reduce the inter-slice spacing at an arbitrary scaling factor without the need for real ground truth.

As a summary, the main contributions in this paper are listed as follows.
\begin{itemize}
    \item We develop a novel SR method named SA-INR for continuous slice interpolation of MR images. With a single model, SA-INR can reduce the inter-slice spacing of MR images at an arbitrary scaling factor, including non-integers.
    \item We design the LASA operation to enhance the spatial awareness of our network in combination with a gradient-guided voxel-wise gating mask. In this manner, our method achieves superior performance while keeping high computational efficiency.
    \item We evaluate the performance of SA-INR on both simulated and real MR images, showing a significant improvement over other state-of-the-art single-scale and arbitrary-scale SR methods. Extensive demos, codes, and models are released via our project website\footnote{Project page: \url{ https://jamesqfreeman.github.io/SA-INR/}}.
\end{itemize}

\section{Related Work}
In this section, we first review the single-image SR methods for natural images and medical images. Then, we explain the idea of implicit neural representation as well as its applications in SR. Finally, we introduce the related work regarding arbitrary-scale SR methods.

\subsection{Single-Image Super-Resolution}
Super-resolution techniques can be categorized into two groups based on the number of images needed for a scene: multiple image super-resolution (MISR) and single image super-resolution (SISR).
Since it is hard to acquire multiple images from different views of the same scene, deep learning based SISR methods are more prevalent and have achieved state-of-the-art performance. 
\citet{SRCNN} used convolutional neural networks (CNN) to learn an end-to-end mapping between the LR and HR images directly and achieved superior performance compared to the traditional methods. Then, a lot of deep learning approaches have been developed and improved by exploring a better network architecture~\citep{VDSR,DRCN,EDSR,RDN,RCAN}. For example, \citet{RDN} proposed a novel residual dense network (RDN) that adopted dense connections at the layer level and block level. 
\citet{EDSR} proposed an enhanced deep super-resolution network (EDSR) that achieved significant improvement by removing unnecessary modules (batch normalization) in residual networks.

Inspired by the progress in natural images, SISR methods are also widely used in the field of medical imaging.
One of the most important applications is to improve the through-plane resolution of MR images.
\citet{DeepResolve} introduced a 3D network of DeepResolve to acquire the high-resolution thin-slice knee MR image from low-resolution thick slices.
To tackle the problem of lack of ground truth, \citet{smore} proposed a self-supervised method to improve the through-plane resolution of MR images. In general, they degraded in-plane slices to prepare 2D HR-LR pairs and trained a regression model between them. Then they applied the regression model in the through-plane direction.
\citet{kai} proposed a two-stage self-supervised framework to reduce the slice spacing of MR images.
In the first stage, HR images of reduced slice spacing were synthesized from real LR images after embedding through a VAE. In the second stage, the synthesized LR-HR image pairs were used to supervise the training of an SR network.
\citet{TSCNet} also introduced a two-stage self-supervised cycle-consistency network (TSCNet) for MR slice interpolation in the axial view. The interpolation network was pre-trained on paired samples in the sagittal and coronal views and then refined by a cyclic procedure involving axial slices.

\subsection{Implicit Neural Representation}
Implicit neural representation (INR) enables us to represent all kinds of signals as a continuous and usually differentiable function.
Most conventional signal representations are discrete, such as the discrete 2D grids of RGB pixels in natural photos and the discrete 3D grids of HU-value voxels in CT. 
The signal parameterized by INR, on the other hand, is implicitly defined, continuous, and differentiable. By using neural networks, the INR signal is parameterized as a continuous function that converts a spatial coordinate to the corresponding signal value.
INR has been widely used, which outperforms grid-, point-, and mesh-based representations in various tasks including 3D shape modeling~\citep{chen2019learning,genova2020local}, volume rendering~\citep{nerf,derf}, and 3D reconstruction~\citep{occupancy,convOccupancy}.

IRN is not restricted by spatial resolution, providing a novel perspective on image representation.
\citet{stanley2007compositional} proposed parameterizing images implicitly via neural networks. 
\citet{sitzmann} replaced ReLU with periodic activation in neural networks to better reconstruct the details.
\citet{bemana2020x} learned an implicit function to output RGB values given the space, time, or light coordinates as input.
\citet{Karras2021} applied the INR in image generation tasks and outperformed its pixel-based competitor.
Recently, \citet{LIIF} introduced the local implicit image function (LIIF) for image SR. Given a coordinate, the function takes the coordinate and queries the local latent code around the coordinate as inputs, and then predicts the RGB value at the given
coordinate as an output. Since the coordinates can be continuously sampled, the output image can be presented in arbitrary spatial resolution.

\subsection{Arbitrary-Scale Super-Resolution}
Arbitrary-scale SR refers to that, for any LR image, the method can continuously zoom in with an arbitrary scaling factor by only using a single model.
There are several related works in the literature.
For example, \citet{MetaSR} adopted the idea of meta-learning and designed an arbitrary-scale MetaSR network. By replacing the traditional upscale module with their proposed Meta-Upscale module, MetaSR is able to generate the scale-specific weights for the filters in the network.
\citet{IREM} proposed to learn a volumetric continuous function for each MR subject based on multiple observed LR images of different views (sagittal, coronal, axial), achieving arbitrary-scale SR using the continuity of the function.
Instead of learning an independent function for each object,
\citet{LIIF} introduced a shared function of LIIF for 2D images, and later \citet{ArSSR} extended it to 3D volumes.
However, ArSSR performs isotropic SR for MR images, which is not common in real clinical settings. 
And in this paper, we will focus on how to perform high-quality through-plane SR, which is more likely to happen in a clinical scenario.

\section{Method}
In this part, we first overview the overall architecture of SA-INR in Section \ref{sec: SA-INR}.
Then, we detail the LASA operation in Section \ref{sec:lasa}. 
Finally, we introduce how the gradient-guided gating mask is developed in Section \ref{sec:mask}.

\begin{figure*}[t]
    \centering
    \includegraphics[width=\textwidth]{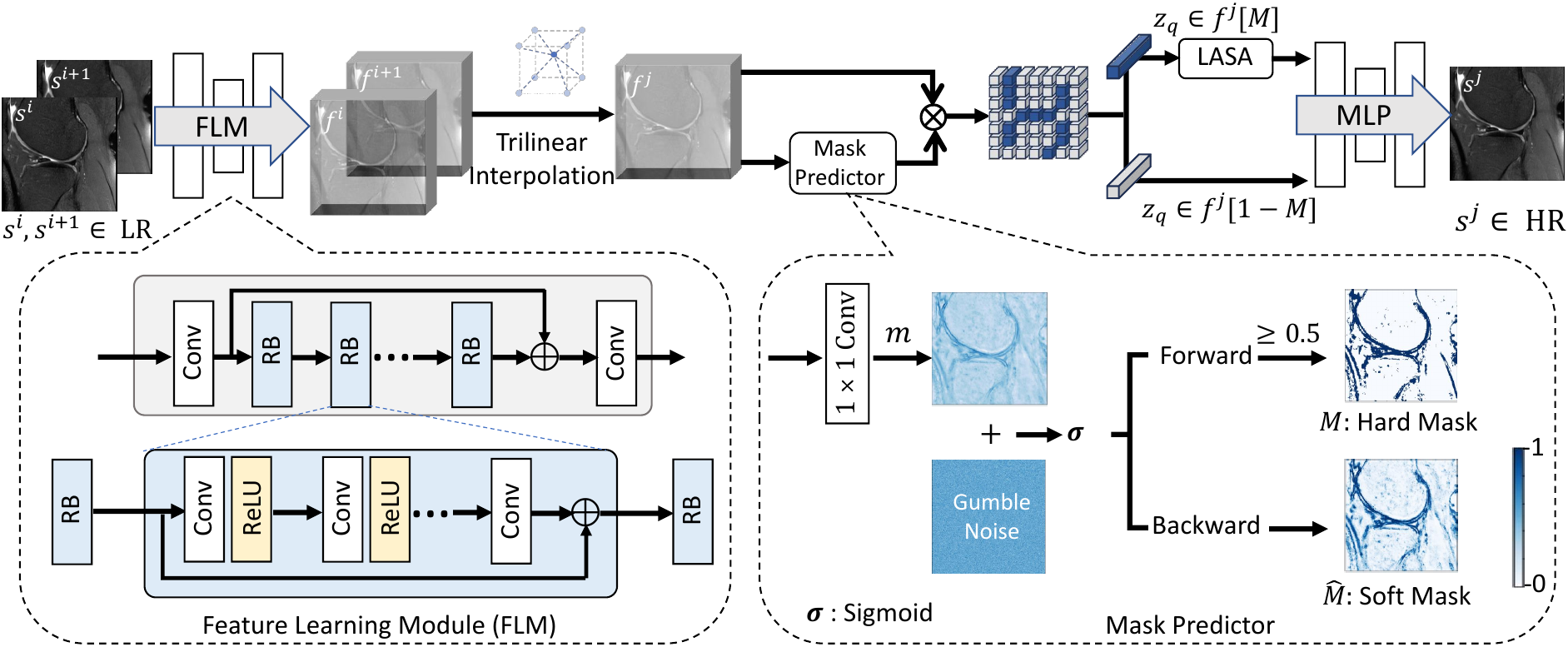}
    \caption{The overall architecture of SA-INR. 
    $s^i$, $s^{i+1}$ are denoted as the two adjacent slices of $s^j$. 
    $f^i$, $f^{i+1}$, and $f^j$ are the feature maps of $s^i$, $s^{i+1}$ and $s^j$, respectively.
    LASA is short for the local-aware spatial attention. 
    $z_q$ is a feature vector sampled from $f^{j}$. $M$ represents a voxel-wise gating mask that specifies the positions to be processed by the LASA operation.
    RB denotes residual block. $m$ is the output of the convolutional layer.}
    \label{fig2}
\end{figure*}

\subsection{SA-INR for Slice Super-Resolution}
\label{sec: SA-INR}
In the scheme of implicit neural representation, an image is perceived as the mapping from the 3D coordinates to the corresponding intensities. The mapping can be parameterized by a decoder such as a multilayer perceptron (MLP). Inspired by that, we propose SA-INR to represent a 3D MR volume as a continuous function, such that for any query coordinate $x_q$, the corresponding voxel intensity $v_q$ can be calculated by:
\begin{gather}
\label{F}
    v_q=F(x_q,z_q).
\end{gather}
The decoding function $F$ takes $x_q$ and the associated implicit representation $z_q$ as input, and outputs the voxel intensity $v_q$ accordingly. Since $x_q$ can be sampled continuously and densely, the HR image with any desired slice spacing can be generated by feeding the corresponding coordinates $\{x_q\}$ and the implicit representations $\{z_q\}$ to the above function. 

The overall architecture of SA-INR is depicted in Fig.~\ref{fig2}. 
Specifically, for each slice in HR, we first find its adjacent slices in LR. Then, we extract the implicit representations for these LR slices.
Next, given the query coordinates $\{x_q\}$ of an HR slice, we generate the corresponding implicit representations $\{z_q\}$ from their neighboring grid voxels in the adjacent slices.
Finally, we decode $\{x_q\}$ concatenated with $\{z_q\}$ to generate the desired HR slice that is between the two adjacent LR slices.

\subsubsection{Slice Localization}
Suppose there are $n_L$ slices in a given LR image with an inter-slice spacing of $c_L$ (unit: mm) and the desired HR image has an inter-slice spacing of $c_{H}$ (unit: mm), then the number of HR slices will be:
\begin{gather}
    n_{H}=\lfloor\frac{(n_{L}-1)*c_{L}}{c_{H}}\rfloor+1.
\end{gather}
$\lfloor\cdot\rfloor$ is the floor function. For the $j$-th slice $s^j\in{\mathbb{R}}^{W\times{H}}$ ($0\leq{j}\leq{n_{H}-1}$) in HR, we could locate the two adjacent slices $s^i$ and $s^{i+1}$ in the LR image by the following projection:
\begin{gather}
    i=\lfloor{j*\frac{c_{H}}{c_{L}}}\rfloor.
\end{gather}
Since the scaling factor $c_{L}/c_{H}$ may not be an integer, all the slices in HR will be reconstructed. 
Note that it is different from many other SR works designed for integer scaling factors, where the original slices in the LR images can be copied directly if they overlap with specific slices in the desired HR. 

\subsubsection{Extraction of Low-Resolution Representation}
After locating two adjacent slices $s^i$ and $s^{i+1}$ of $s^j$, we input them into the feature learning module to learn an implicit representation for each grid voxel in $s^i$ and $s^{i+1}$. 
The feature learning module then outputs two feature maps $f^i\in{\mathbb{R}}^{W\times{H\times{C}}}$ and $f^{i+1}\in{\mathbb{R}}^{W\times{H\times{C}}}$, respectively.
Each $C$-dim feature vector on the feature maps is regarded as an implicit representation, preserving local appearance information of the corresponding grid coordinate,
as well as the implicit spatial relationship with nearby coordinates.

In particular, we adopt the Enhanced Deep Super-Resolution Network (EDSR)~\citep{EDSR} as the backbone of our feature learning module. As shown in the bottom-left of Fig.~\ref{fig2}, the module comprises several residual blocks. Each block contains two convolutional layers with a skip connection. No normalization layer is used. The outputs from the first convolutional layer and the last residual block are combined through a long-range skip connection.

\begin{figure*}[t]
    \centering
    \includegraphics[width=\textwidth]{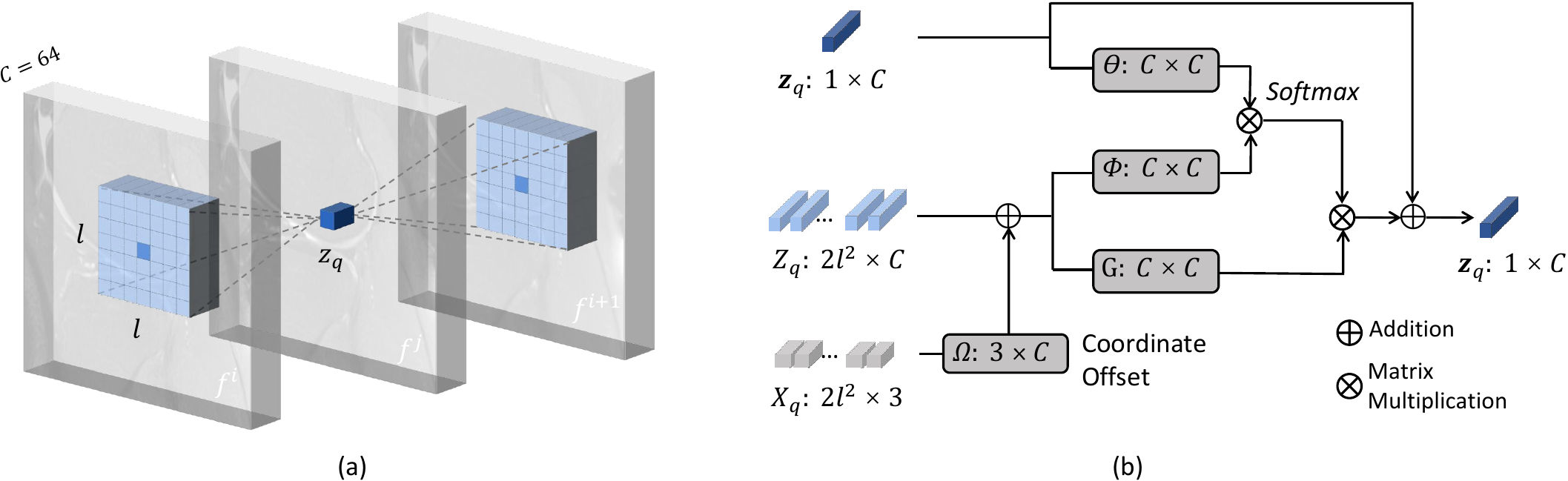}
    \caption{
    (a) The illustration of the proposed local-aware spatial attention (LASA) operation.
    (b) The diagram to compute the local-aware attention. 
     ${\Theta}$, $\Phi$ and $\mathrm{G}$ denote the weights of the linear embedding layers. $\Omega$ denotes the weight of the coordinate embedding layer. 
     The dark blue tensor ($\boldsymbol{z}_q$) denotes the implicit representation at the position of $\boldsymbol{x}_q$. 
     The light blue tensors ($Z_q$) denote the neighborhood of $\boldsymbol{z}_q$. The grey tensors ($X_q$) represent the 3D coordinate offsets between $\boldsymbol{z}_q$ and its neighbors. 
    }
    \label{fig4}
\end{figure*}


\subsubsection{Estimation of High-Resolution Representation}
Next, we aim to derive the feature map $f^{j}$ of $s^{j}$ from the extracted $f^i$ and $f^{i+1}$.
The simple trilinear interpolation is adopted to get an initial estimation of $f^j$, though its accuracy is restricted due to the combination of nearby voxels with fixed weights and the limited receptive field.  
To address these issues, we propose the Local-Aware Spatial Attention (LASA) operation to refine the feature vector in $f^{j}$ by capturing dynamic weights with the neighbors in $f^i$ and $f^{i+1}$.
Also, LASA allows for a larger interpolation receptive field, thus the feature vector in $f^{j}$ can acquire more information from a larger neighborhood in the adjacent LR slices.
The details of the LASA operation are introduced in Section \ref{sec:lasa}.

Although performing the LASA operation for each feature vector on $f^j$ can significantly improve SR performance, we find it comes at a high computational cost.
Therefore, it is more practical to only apply the LASA operation to the areas where conventional SR may perform poorly, such as near the edges of the tissues or areas with rapidly changing intensity values. 
For this purpose, we design a gradient-guided gating mask to help determine the locations that require LASA or not.
In this way, we could achieve comparable and even better performance while reducing the expensive computational cost incurred by the attention operation.

As can be seen in Fig.~\ref{fig2}, we input the initial $f^{j}$, which is generated from trilinear interpolation, to the mask predictor and to generate the gating mask $M$.
The LASA operation is only applied to update the feature vectors identified necessary by $M$ (represented as blue tensors in Fig.~\ref{fig2}). 
To derive the mask predictor, 1$\times$1 convolution is used in combination with the gumbel-softmax trick~\citep{gumbel}.
The details of the gradient-guided gating mask are introduced in Section \ref{sec:mask}.

\subsubsection{High-Resolution Decoding}
Finally, for each grid voxel on $s^j$, we feed its 3D coordinate $x_q$ and the feature vector $z_q$ into the function $F$ in \eqref{F}, which then returns the corresponding intensity value. 
The function $F$ is parameterized as a 5-layer MLP with ReLU activation, where the dimension of hidden layers is set to 256. 
The whole slice $s^{j}$ is generated by iterating over all of its grid voxels.

Note that we adopt the way of residual learning as in \citet{VDSR}. Let $s^{j}_{\mathrm{img}}$ denote the result of trilinear interpolation in the image space. We combine $s^{j}_{\mathrm{img}}$ with the model output $s^{j}_{\mathrm{model}}$ to form the predicted final result:
\begin{gather}
    s^{j}_{\mathrm{pred}}=s^{j}_{\mathrm{img}}+s^{j}_{\mathrm{model}}.
\end{gather}
The goal of the SA-INR network is to minimize the difference between the predicted image and the ground-truth image,
which is evaluated by the ${L_1}$ loss:
\begin{gather}
    L_\mathrm{fidelity}=|s^{j}_{\mathrm{pred}}-s^{j}_{\mathrm{gt}}|.
\end{gather}

\subsection{Local-Aware Spatial Attention}
\label{sec:lasa}
To better model the spatial dependency between each query coordinate and its neighbors, the LASA operation is designed. 
Given a learned mask $M$ by the mask predictor, we conditionally apply the proposed LASA operation for $z_q\in{f^{j}[M]}$.
In order to suppress irrelevant features and reduce computational burdens, we limit the calculation of attention to a certain neighborhood.
In Fig.~\ref{fig4}, for a $\boldsymbol{z}_q$ on $f^j$, its neighborhood is defined to cover an $l\times l$ window ($l$=7 in our implementation) on feature map $f^i$, and a corresponding window on feature map $f^{i+1}$. 
Let $\mathrm{Z}_q\in{\mathbb{R}^{2l^2 \times C}}$ denote the flattened feature vectors in the neighborhood of $\boldsymbol{z}_q$, then we update ${\boldsymbol{z}}_q$ as
\begin{gather}
\centering
    {\boldsymbol{z}}_q \leftarrow \text{softmax}(\boldsymbol{z}_q{\Theta}(\mathrm{Z}_q{\Phi})^T)\mathrm{Z}_q{\mathrm{G}}+\boldsymbol{z}_q,
\end{gather}
where 
${\Theta}$, $\Phi$ and $\mathrm{G}$ denote the learnable weight matrices of the linear layers. 
We use dot-product to compute the similarity between $\boldsymbol{z}_q$ and its neighbors in $\mathrm{Z}_q$.
Note that we follow \citet{NL}, adding input to the output of the attention.

Also, we use the coordinate offset information to help estimate $\boldsymbol{z}_q$ more accurately:
\begin{equation}
      {\boldsymbol{z}}_q \leftarrow \text{softmax}(\boldsymbol{z}_q{\Theta}((\mathrm{Z}_q+\mathrm{X}_{q}{\Omega}){\Phi})^T)(\mathrm{Z}_q+\mathrm{X}_{q}{\Omega}){\mathrm{G}}+\boldsymbol{z}_q,
\end{equation}
where $\mathrm{X}_q$ denotes the coordinate offsets between $\boldsymbol{z}_q$ and its neighbors.
Specifically, an embedding layer $\Omega$ is employed to map the three-dimensional coordinate offset to $C$-dimensional feature. 

\subsection{Gradient-Guided Gating Mask}
\label{sec:mask}
\begin{figure}[t]
    \centering
    \includegraphics[width=0.35\textwidth,height=0.35\textwidth]{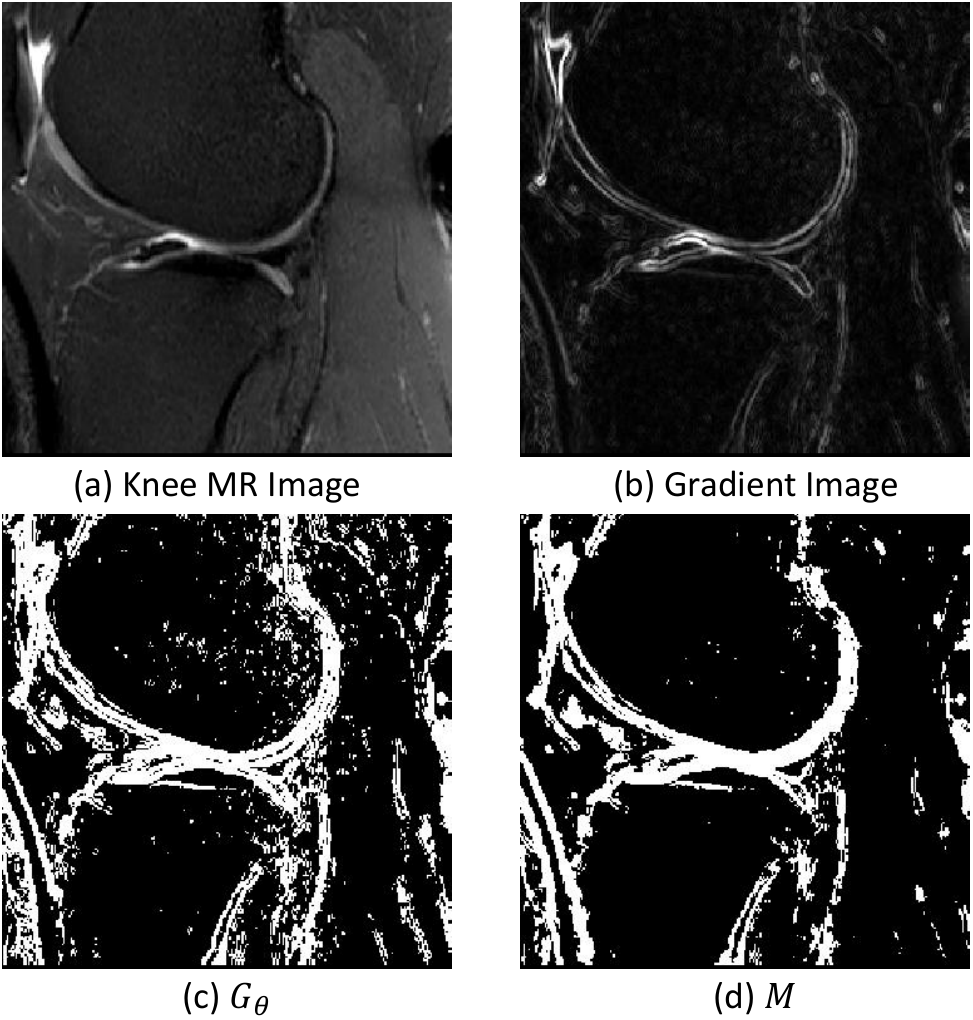}
    \caption{The illustration of gradient-guided gating mask on a knee MR image. (a) The input knee MR image. (b) The corresponding gradient map. (c) $G_\theta$ denotes the binary mask via thresholding of the gradient image. (d) $M$ denotes the binary mask predicted by our model.}
    \label{fig3}
\end{figure}

The LASA operation allows each query coordinate to incorporate more information from a larger receptive field. 
However, executing LASA for each query coordinate will incur a significant computational overhead. 
A feasible way is to conditionally apply LASA based on the complexity of the area under consideration. 
Given a knee MR image in Fig.~\ref{fig3}(a) as an example, the corresponding intensity gradient map is calculated and displayed in Fig.~\ref{fig3}(b). 
We argue that it is more difficult to process SR for those areas of high gradient, e.g., near the edges of tissues or organs. 
Therefore, we could use simple thresholding to obtain a binary gating mask $G_\theta$ and conditionally apply LASA according to $G_\theta$.
In the example of Fig.~\ref{fig3}(c), the threshold is set as $20\%$ quantile of the gradient map, so that only top-20$\%$ positions are selected for the expensive attention operation. 

However, the resulting mask after thresholding tends to be noisy and rough, as seen in Fig.~\ref{fig3}(c).
Besides, in the inference stage for a real SR task, the HR slice is not available in the beginning; thus there is no corresponding gradient map. 
Therefore, we propose to learn a mask that specifies the positions to which LASA should be applied.
In Fig.~\ref{fig3}(d), we display the learned mask $M$ produced by our method. 
Compared to $G_\theta$, the two masks consume similar computation, yet the learned mask $M$ is less affected by many isolated noises as in $G_\theta$. 

To learn the gating mask, in Fig. \ref{fig2}, we set a 1 $\times$ 1 convolution layer as our mask predictor.
Given an input $f^j\in{\mathbb{R}}^{W\times{H\times{C}}}$, the mask predictor outputs $m\in{\mathbb{R}}^{W\times{H}}$.
Then we derive a soft gating mask $\hat{M}$ by
\begin{gather}
    \hat{M}=\sigma(m),
\end{gather}
where $\sigma$ denotes the sigmoid function, and $\hat{M}\in{\mathbb{R}}^{W\times{H}}$ represents the probability that a feature vector should be processed by LASA.
Correspondingly, $(1-\hat{M})$ indicates the probability of a feature vector not being processed.

%
In the training stage, $M$ can be randomly sampled from the Bernoulli distribution defined by $\hat{M}$.
To address the issue that the sampling process is not derivable, the reparameterization strategy \citet{vae,chang2019differentiable,herrmann2020channel,Verelst_2020} is widely investigated.
Similar to~\citep{Verelst_2020}, we adopt binary gumbel-softmax to approximate the generation of the binary gating mask, where the gumble noise is introduced:
\begin{gather}
    \hat{M}=\sigma({m+g_1-g_2}).
\end{gather}
Here $g_1$ and $g_2$ are two random variables drawn from the gumble distribution~\citep{gumbel} controlled by $\mu$ and $\beta$:
\begin{gather}
    f(g;\mu,\beta)=e^{-z-e^{-z}},\;z=\frac{g-\mu}{\beta}.
\end{gather}

Note that the gumble noise is introduced in the training stage only.
In the forward pass of training (as well as the inference stage), we can acquire a hard gating mask $M$ by:
\begin{gather}
    \label{eq-9}
    M= \begin{cases}
    1 \qquad \hat{M}\ge0.5;\\
    0 \qquad \hat{M}<0.5.
    \end{cases}
\end{gather}
While in the backward pass, we set $M=\hat{M}$ for gradient backpropagation.

The percentile of the activated positions in the mask is critical to control the computation cost. 
In Fig.~\ref{fig3}(c), we show $\theta\%$ (=20$\%$) positions are potentially considered after applying thresholding to the gradient map.  
Similarly, to derive a learnable mask, we can count the proportion of the activated positions in $M$ as
\begin{gather}
    \hat{\theta}=\mathrm{sum} (M)/(W*H),
\end{gather}
and require $\hat{\theta}$ to be close to our budget $\theta$.

In summary, we define the mask loss as 
\begin{gather}
    L_{\mathrm{mask}}=\gamma\rm{CE}(G_\theta,M)+(\theta-\hat{\theta})^2,
\end{gather}
where CE denotes the cross entropy loss such that the learned mask $M$ can roughly follow $G_\theta$ and focus more on the difficult areas.
Since $G_\theta$ is sensitive to noise, we use it only in the early stage of training for better convergence.
Specifically, we use $\gamma$ to control the weight of $G_\theta$ and gradually reduce it to zero during training.

Finally, the whole SA-INR network in Fig. \ref{fig2} is optimized by the following overall loss:
\begin{gather}
    L=L_{\mathrm{fidelity}} + L_{\mathrm{mask}}.
\end{gather}

\section{Experiments}
In order to thoroughly evaluate our proposed method, we have designed three sets of experiments.
\begin{enumerate}
    \item \textbf{Experiments on SR of Brain MRI.} To quantitatively evaluate the performance, we conduct simulation experiments with 1,113 isotropic T1-weighted brain MR images from the HCP-1200 dataset~\citep{HCP}. 
    We treat those images as HR ground truth, from which we simulate the corresponding LR images by randomly discarding slices at a stride of 2, 3, and 4. 
    The simulated pairs of HR and LR images are used to train and validate the SR models under comparison. 
    Although we can only simulate LR-HR pairs of limited integer scaling factors, it is worth noting that our trained SA-INR network can produce images for any non-integer factor.

    \item  \textbf{Experiments on Downstream Tissue Segmentation of Brain MRI.} To verify the effectiveness of our method on downstream tasks such as tissue segmentation, we adopt Fastsurfer~\citep{fastsurfer}, a deep-learning-based open-source tool, to perform brain tissue segmentation on the SR results from different methods. 
    The quality of the SR images can be further validated by analyzing the differences between the segmentation results of the SR images and the ground-truth images qualitatively and quantitatively.
    
    \item  \textbf{Experiments on SR of Clinical Knee MRI.} To demonstrate the real-case usage of the proposed method, we conduct experiments on clinical knee MR images following our previous work~\citep{kai}. Specifically, we have collected 2,627 knee MR images, whose intra-slice resolution is 0.3mm$\times$0.3mm and inter-slice spacing is 3.3mm or 4.5mm. Since no HR ground truth is available here, we adopt the previously proposed two-stage framework for self-supervised super-resolution to mimic the real clinical scenario.

\end{enumerate}

\begin{table*}
\centering
\caption{The mean and standard deviation of PSNR/SSIM for the scaling factor $\times$2, $\times$3 and $\times$4 on the HCP dataset.}\label{tab1}
\begin{tabular}{c |c c |c c |c c}
    \hline
\multirow{2}{*}{Method} &\multicolumn{2}{c}{$\times$2} &\multicolumn{2}{|c}{$\times$3} &\multicolumn{2}{|c}{$\times$4}\\
\cline{2-7}
\multirow{2}{*}{}&PSNR &SSIM &PSNR &SSIM &PSNR &SSIM\\
\hline
Interpolation &40.59$\pm$2.42 &0.9908$\pm$0.0029 &37.56$\pm$2.41 &0.9821$\pm$0.0054 &36.05$\pm$2.39 &0.9758$\pm$0.0070\\
\hline
EDSR ($\times$2) &43.50$\pm$2.40 &0.9946$\pm$0.0019 &39.51$\pm$2.39 &0.9875$\pm$0.0042 &37.26$\pm$2.29 &0.9793$\pm$0.0071\\
\hline
EDSR ($\times$3) &38.45$\pm$2.53 &0.9815$\pm$0.0071 &41.29$\pm$2.37 &0.9915$\pm$0.0029 &37.16$\pm$2.43 &0.9722$\pm$0.0104\\
\hline
EDSR ($\times$4) &35.02$\pm$2.53 &0.9615$\pm$0.0125 &38.57$\pm$2.45 &0.9786$\pm$0.0084 &39.53$\pm$2.35 &0.9880$\pm$0.0039\\
\hline
MetaSR &43.15$\pm$2.35 &0.9943$\pm$0.0020 &40.86$\pm$2.31 &0.9908$\pm$0.0030 &39.30$\pm$2.29 &0.9876$\pm$0.0039\\
\hline
ArSSR &42.59$\pm$2.33 &0.9936$\pm$0.0022 &40.71$\pm$2.31 &0.9903$\pm$0.0032 &39.65$\pm$2.28 &{0.9884$\pm$0.0037}\\
\hline
Proposed &\cellcolor{black!25}43.62$\pm$2.40 &\cellcolor{black!25}0.9948$\pm$0.0018 & \cellcolor{black!25}41.50$\pm$2.36 & \cellcolor{black!25}0.9919$\pm$0.0028 & \cellcolor{black!25}40.16$\pm$2.33 & \cellcolor{black!25}0.9895$\pm$0.0035\\
\hline
\end{tabular}
\end{table*}
\begin{table*}
\centering
\caption{The sharpness metrics (S3) of three views for the scaling factor $\times$2, $\times$3 and $\times$4 on the HCP dataset.}\label{tab2}

\begin{tabular}{c|ccc|ccc|ccc}
\hline
\multirow{2}{*}{Method} & \multicolumn{3}{c|}{$\times$2} & \multicolumn{3}{c|}{$\times$3} & \multicolumn{3}{c}{$\times$4}                                      \\ \cline{2-10} 
& \multicolumn{1}{c}{Axial}  & \multicolumn{1}{c}{Coronal} & Sagittal & \multicolumn{1}{c}{Axial}  & \multicolumn{1}{c}{Coronal} & Sagittal & \multicolumn{1}{c}{Axial}  & \multicolumn{1}{c}{Coronal} & Sagittal 
\\ \hline
Interpolation              & \multicolumn{1}{c}{0.2196} & \multicolumn{1}{c}{0.2824}  & 0.2779   & \multicolumn{1}{c}{0.2086} & \multicolumn{1}{c}{0.2677}  & 0.2711   & \multicolumn{1}{c}{0.2025} & \multicolumn{1}{c}{0.2598}  & 0.2702   
\\ \hline
EDSR                       & \multicolumn{1}{c}{0.2300} & \multicolumn{1}{c}{0.2973}  & 0.2912   & \multicolumn{1}{c}{0.2231} & \multicolumn{1}{c}{0.2882}  & 0.2772   & \multicolumn{1}{c}{0.2168} & \multicolumn{1}{c}{0.2800}  & 0.2690   
\\ \hline
MetaSR                     & \multicolumn{1}{c}{0.2292} & \multicolumn{1}{c}{0.2963}  & 0.2826   & \multicolumn{1}{c}{0.2237} & \multicolumn{1}{c}{0.2890}  & 0.2738   & \multicolumn{1}{c}{0.2175} & \multicolumn{1}{c}{0.2807}  & 0.2682   
\\ \hline
ArSSR                      & \multicolumn{1}{c}{0.2313} & \multicolumn{1}{c}{0.2993}  & \cellcolor{black!25}0.2919   & \multicolumn{1}{c}{0.2249} & \multicolumn{1}{c}{0.2910}  & 0.2766   & \multicolumn{1}{c}{0.2197} & \multicolumn{1}{c}{0.2844}  & 0.2778   
\\ \hline
Proposed                   & \cellcolor{black!25}0.2331 & \cellcolor{black!25}0.3017  & 0.2881   & \cellcolor{black!25}0.2287 & \cellcolor{black!25}0.2960  & \cellcolor{black!25}0.2813   & \cellcolor{black!25}0.2244 & \cellcolor{black!25}0.2907  & \cellcolor{black!25}0.2796   \\ \hline
\end{tabular}
\end{table*}
\begin{table*}[ht]
\centering
\caption{Ablation Study on LASA and Gating Mask.}
\begin{tabular}{cc|ccc|ccc|ccc}
\hline
\multicolumn{2}{c|}{Method} & \multicolumn{3}{c|}{$\times$2} & \multicolumn{3}{c|}{$\times$3} & \multicolumn{3}{c}{$\times$4}  \\ \hline
  LASA &      Gating Mask                & PSNR  & SSIM   & GFLOPs & PSNR  & SSIM   & GFLOPs & PSNR  & SSIM   & GFLOPs \\ \hline
    -- & --          & 42.96 & 0.9940 & 288.6  & 41.20 & 0.9914 & 302.6  & 39.78 & 0.9884 & 316.5  \\ \hline
   \Checkmark& --           & 43.62 & 0.9948 & 447.3  & 41.50 & 0.9919 & 514.2  & 40.16 & 0.9895 & 580.3  \\ \hline
 \Checkmark &  \Checkmark            & 43.66 & 0.9949 & 320.3  & 41.54 & 0.9920 & 344.9  & 40.21 & 0.9896 & 369.4  \\ \hline
\end{tabular}
\label{tab:ablation}
\end{table*}

\begin{figure*}[t]
    \begin{center}
    \includegraphics[width=1\textwidth]{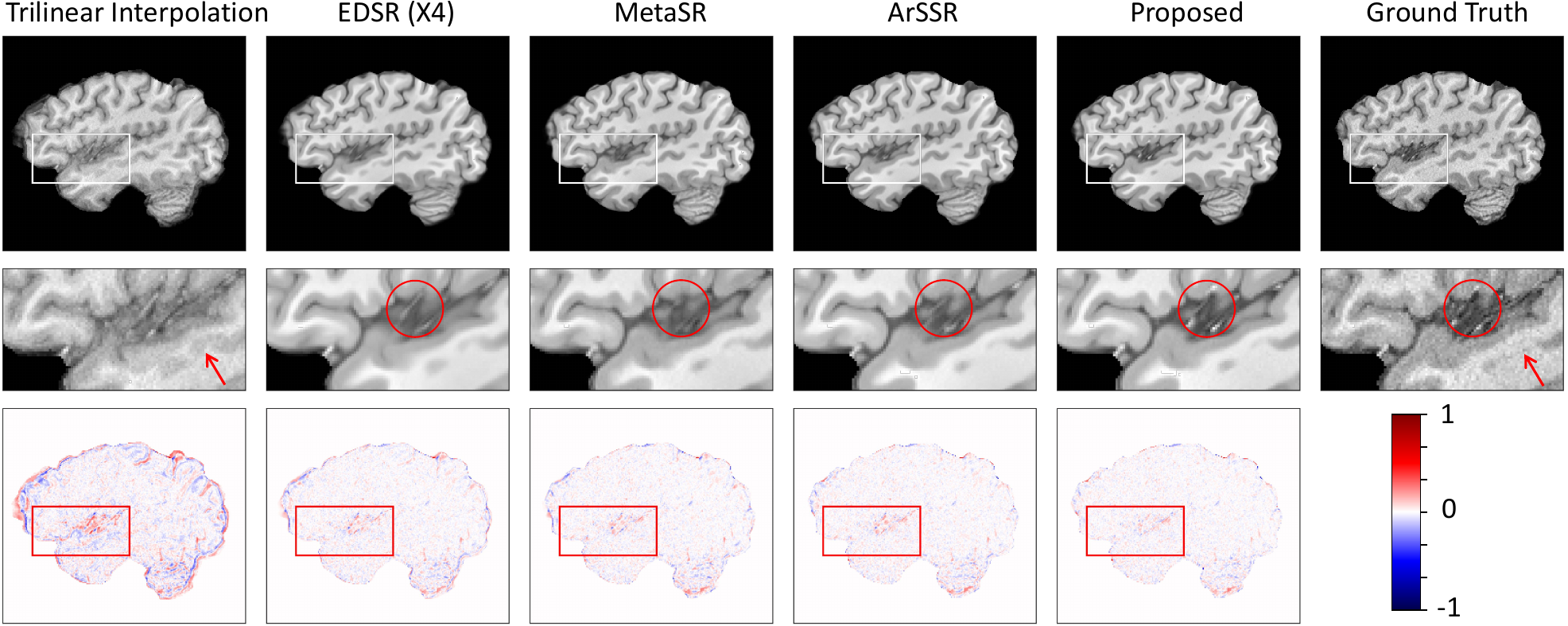}
    \caption{Visual comparisons of a $\times4$ test case from HCP-1200 dataset by all the comparing methods in the sagittal (in-plane) view.
Each column displays the slice constructed by different methods, while the last column indicates the ground truth.
The error maps of individual results with respect to the ground truth are shown in the last row.} \label{fig5}
    \end{center}
\end{figure*}


\subsection{Experimental Settings}
\subsubsection{Comparison Methods}
To evaluate the performance of our proposed method, we compare it with several existing methods, including both single-scale and arbitrary-scale SR methods.
\begin{itemize}
\item Trilinear Interpolation: a multivariate linear interpolation method for 3D images, which is widely adopted in the field of medical imaging.
\item EDSR $(\times{k})$~\citep{EDSR}: a deep-learning-based SR network trained at a single factor of $k$, which is often regarded as the baseline for single-scale SR in deep learning.
\item MetaSR~\citep{MetaSR}: a meta-learning-based arbitrary-scale SR method, which can dynamically generate the parameters of the SR network for each specific scaling factor.
\item ArSSR~\citep{ArSSR}: a straightforward 3D implementation of LIIF, which learns a continuous voxel representation upon the 3D coordinates for arbitrary-scale SR.
\end{itemize}

Since the original EDSR and MetaSR are designed for 2D images, we reimplement these methods by extending them to 3D.  
In detail, we replace the 2D convolutional kernels of EDSR with 3D ones and use trilinear interpolation for upsampling. 
Besides, we extend three important modules in MetaSR, namely Location Projection, Weight Prediction, and Feature Mapping, to 3D. For example, Weight Prediction uses a network to predict the weights of the 2D filters for natural images, while we change the dimension of the weights to three. 
As for ArSSR, the original code can only carry out SR for isotropic scaling factors ($k\times k\times k$). We modify it to adapt to the scenario of reducing slice spacing ($1\times 1\times k$). 
And for a fair comparison, the feature learning modules in ArSSR and MetaSR follow the EDSR architecture, which is the same as our proposed method.

\subsubsection{Comparison Metrics}
For quantitative comparisons, we use several popular metrics to evaluate the performance of the SR methods. 
The first metric is the peak signal-to-noise ratio (PSNR)~\citep{PSNR}, which is defined as
\begin{gather}
    {\rm PSNR}=10\cdot{\log_{10}({\frac{{\rm MAX}^2}{\rm MSE}})}.
\end{gather}
Here, MAX is the maximum intensity value of the ground-truth image. MSE denotes the mean squared error between the SR image and the ground truth.

The second metric is the structural similarity index (SSIM)~\citep{SSIM}, which is calculated as follows:
\begin{gather}
    {\rm SSIM}=\frac{(2\mu_{sr}\mu_{gt}+c_1)(2\sigma_{{sr},{gt}}+c_2)}{(\mu_{sr}^2+\mu_{gt}^2+c_1)(\sigma_{sr}^2+\sigma_{gt}^2+c_2)}.
\end{gather}
Here $\mu_{sr}$ and $\mu_{gt}$ are the average voxel intensities in $I_{\rm sr}$ and $I_{\rm gt}$, ${\sigma_{sr}}$ and $\sigma_{gt}$ are the variances, $\sigma_{{sr},{gt}}$ is the covariance. And $c_1$ and $c_2$ are used to stabilize the possibly weak denominator.

Although PSNR and SSIM are widely used to evaluate image quality, both of them can be biased toward blurred images. 
Therefore, in addition to PSNR and SSIM, we add an extra evaluation metric of S3~\citep{s3} to quantify the locally perceived sharpness of an image. 
For a better perception of sharpness, S3 combines a spectral measure based on the slope of the local magnitude spectrum and a spatial measure based on local maximum total variation (TV).

\subsubsection{Implementation Detail}
\label{patch_training}
We provide the details of our implementation below.

\textbf{Training}. 
We adopt patch-style training to save memory costs.
Let $k$ be the scaling factor. 
In each training iteration, we randomly draw a $k\in\{1,2,3,4\}$ and randomly crop an HR patch with a size of $64\times64\times(16k+1)$ from an HR image.
The LR patch with a size of $64\times64\times17$ is simulated by discarding the slices from the cropped HR patch at a stride of $k$. The resulting LR patch is paired with the HR patch with a slice spacing ratio of $k$.

The training process then follows the four steps introduced in the method part, which is supervised by the previous LR-HR pairs.
First, for each $x_q=(x,y,z)$ in HR patch, we locate its neighbors in LR. 
Then, we acquire the LR feature maps.
Next, we estimate the implicit representation $z_q$ for each $x_q=(x,y,z)$. 
Finally, we concatenate $\{x_q\}$ and $\{z_q\}$ and input them into the decoding function to acquire the final result.

\textbf{Inference}. Since inference is a derivative-free process, it is practical to input the whole LR image into the network. 
Specifically, we first input the whole LR image into the feature learning module and get the corresponding feature maps. 
Then, we query the corresponding implicit representations $\{z_q\}$ for coordinates $\{x_q\}$ within each slice in HR. 
Next, we feed $\{x_q\}$ and $\{z_q\}$ to the decoding function. The final HR image is generated slice by slice.

\textbf{Experimental Environment}. Our network is implemented with PyTorch~\citep{pytorch} and trained for 2000 epochs. The training of our model takes about 100 GPU hours with NVIDIA RTX 3090 24GB. Adam optimizer~\citep{Adam} is used with a learning rate of $1e^{-4}$. The computational budget $\theta$ for the mask predictor is set to 20\% in both brain and knee experiments.
At the start of training, we set $\gamma$ to 1 and reduce it to half every 50 epochs.

\begin{table*}[]
\centering
\caption{ Quantitative results (Dice) of the fully automatic segmentation for scaling factor $\times$2, $\times$3 and $\times$4 on the HCP dataset.}\label{tab:seg}
\begin{tabular}{c|cp{1.55cm}<{\centering}p{1.55cm}<{\centering}p{1.55cm}<{\centering}p{1.55cm}<{\centering}p{1.55cm}<{\centering}p{1.55cm}<{\centering}p{1.55cm}<{\centering}c}
\hline
\begin{tabular}[c]{@{}c@{}}Scaling \\ Factor\end{tabular}               & Method        & \begin{tabular}[c]{@{}c@{}}Cortical \\ White Matter\end{tabular} & \begin{tabular}[c]{@{}c@{}}Cortical \\ Gray Matter\end{tabular} & \begin{tabular}[c]{@{}c@{}}Cerebellar \\ White Matter\end{tabular} & \begin{tabular}[c]{@{}c@{}}Cerebellar \\ Cortex\end{tabular} & \begin{tabular}[c]{@{}c@{}}Brain \\ Stem\end{tabular} & CSF    & Overall &  \\
\hline
\multirow{5}{*}{$\times$2} & Interpolation & 0.9760  & 0.9386  & 0.9698  & 0.9735 & 0.9840  & 0.9369 & 0.9402  &  \\
                    & EDSR($\times$2)      & 0.9832  & 0.9549 & 0.9744  & 0.9773   & 0.9870  & 0.9583 & 0.9572  &  \\
                    & MetaSR        & 0.9822  & 0.9517 & 0.9728  & 0.9757 & 0.9863    & 0.9534 & 0.9539  &  \\
                    & ArSSR         & 0.9830   & 0.9544  & 0.9737  & 0.9772 & 0.9869 & 0.9560 & 0.9565  &  \\
                    & Proposed      &    \cellcolor{black!25}0.9840	 & \cellcolor{black!25}0.9573  & \cellcolor{black!25}0.9750 & \cellcolor{black!25}0.9787  & \cellcolor{black!25}0.9874  & \cellcolor{black!25}0.9583 &  \cellcolor{black!25}0.9593  &  \\
\hline
\multirow{5}{*}{$\times$3} & Interpolation & 0.9557                                                           & 0.8904                                                          & 0.9450                                                             & 0.9550                                                       & 0.9718     & 0.8637 & 0.8907  &  \\
                    & EDSR($\times$3)      & 0.9769                                                           & 0.9386                                                          & 0.9618                                                             & 0.9674                                                       & 0.9813     & 0.9298 & 0.9401  &  \\
                    & MetaSR        & 0.9751                                                           & 0.9344                                                          & 0.9585                                                             & 0.9649                                                       & 0.9795     & 0.9098 & 0.9343  &  \\
                    & ArSSR         & 0.9733                                                           & 0.9309                                                          & 0.9585                                                             & 0.9645                                                       & 0.9804     & 0.9227 & 0.9329  &  \\
                    & Proposed      & \cellcolor{black!25}0.9777                                                                &  	\cellcolor{black!25}0.9411                                    &             	\cellcolor{black!25}0.9631	                                                                               &    \cellcolor{black!25}0.9691                   &   	\cellcolor{black!25}0.9821            &      	\cellcolor{black!25}0.9324  &   	\cellcolor{black!25}0.9426       &  \\
\hline
\multirow{5}{*}{$\times$4} & Interpolation & 0.9112                                                           & 0.7559                                                          & 0.9089                                                             & 0.8435                                                       & 0.9392     & 0.8148 & 0.7680  &  \\
                    & EDSR($\times$4)      & 0.9709                                                           & 0.9233                                                          & 0.9533                                                             & 0.9598                                                       & 0.9765     & 0.9046 & 0.9239  &  \\
                    & MetaSR        & 0.9672                                                           & 0.9152                                                          & 0.9470                                                             & 0.9553                                                       & 0.9782     & 0.9645 & 0.9131  &  \\
                    & ArSSR         & 0.9687                                                           & 0.9171                                                          & 0.9483                                                             & 0.9554                                                       & 0.9746     & 0.8903 & 0.9176  &  \\
                    & Proposed      & \cellcolor{black!25}0.9729                                                           & \cellcolor{black!25}0.9289                                                       & \cellcolor{black!25}0.9556                                                            & \cellcolor{black!25}0.9624                                                   & \cellcolor{black!25}0.9777     & \cellcolor{black!25}0.9107 &  \cellcolor{black!25}0.9295  & \\
\hline
\end{tabular}
\end{table*}

\subsection{Simulation Experiments on Brain MRI }
\label{sec:sup_brain}

\subsubsection{Dataset and Preprocessing}
We have collected 1,113 subjects of 3T MR images from the HCP-1200 dataset~\citep{HCP}. 
All the MR images have an isotropic voxel spacing of 0.7mm$\times$0.7mm$\times$0.7mm. 
Among them, 891 images are used for training, and 222 images are for testing.
For preprocessing, we apply N4 bias correction and skull-stripping. Note that skull-stripping is required here, as the face/skull is intentionally blurred in HCP to protect the privacy of the subjects.
To simulate the LR images with large slice spacing, the isotropic HR volumes are downsampled in perpendicular to the sagittal view.

\subsubsection{Quantitative Comparisons} 
The quantitative results (PSNR and SSIM) of different algorithms are shown in Table~\ref{tab1}. 
At all scaling factors, our model outperforms other methods (\textit{p}$<$0.01 with paired \textit{t}-tests).

We can find that EDSR ($\times$2) achieves a good result in the task of reducing slice spacing by $k$=2. However, when applying this model to other factors such as $k$=3 or $4$, EDSR ($\times$2) has relatively poor performance. 
Similar findings are also observed when applying EDSR ($\times$3) and EDSR ($\times$4) to the SR tasks of mismatching scaling factors. These results confirm that the conventional SR methods are dedicated to fixed scaling factors and cannot be flexibly generalized to other factors. 

On the contrary, SA-INR can handle multi-factor SR tasks adaptively, even though only a single model is trained. 
For example, our method achieves 43.62dB on PSNR for $\times$2, which is higher than EDSR ($\times$2) with 43.50dB. 
And when applied to $\times$3 and $\times$4, our method has shown more advantages.

Furthermore, compared with MetaSR and ArSSR, which also support arbitrary SR factors, our proposed method leads in PSNR and SSIM at all scaling factors. 
Similar to the two compared methods, our method weighs local neighbors to derive the implicit representation of the query coordinate. 
However, our method differs from them a lot in designing the local receptive field and calculating the weights. 
MetaSR uses a meta-learning module to predict the weights of local $3^3$ neighbors. 
ArSSR relies on the coordinate distance to decide the weights of local $2^3$ neighbors.
Our method effectively enlarges the receptive field (\ie, $2\times7^2$ in our implementation) and enables each query coordinate to learn dynamic weights by attending to its neighbors.

In addition, we compare the sharpness of the images generated by different methods. We compute the average S3 scores for the axial, coronal and sagittal views, respectively. 
The results are shown in Table~\ref{tab2}, where the EDSR models are trained and inferred at the same scaling factors only.
One can notice that our proposed model outperforms other methods on most views and tasks of different scaling factors. 

\subsubsection{Visual Comparisons} 
For visual inspection, we provide an example for the $\times4$ task. 
The input LR image is simulated from the HR image by removing the middle three slices from every five slices.
We then reconstruct all five slices, including the first and the last slices that overlap with the input.
The visual results are shown in Fig.~\ref{fig5}. 
The first row shows the sagittal slices, and the second row provides zoomed-in views.
In the third row, we calculate the error maps compared to the isotropic ground truth.
As pointed out by the red arrows, trilinear interpolation has produced artifacts near the grey matter by simply computing the missing slice as the weighted average of adjacent slices.
Also, by looking at the regions in the red circles, one can notice that EDSR, MetaSR, and ArSSR all fail to reconstruct the high-intensity signals existing in the ground truth. 
In comparison, our proposed SA-INR can recover these signals better. 
Meanwhile, we observe that the noise in the constructed slices tends to be reduced, which is caused by the denoising and smoothing effect often accompanying SR.


\subsubsection{Computation Efficiency} 
To evaluate the efficiency of the proposed model, we have designed an ablation study by removing LASA or the gating mask to observe the differences in image quality and computational cost.
The image quality is measured by PSNR and SSIM between the SR image and the ground truth.
And the computational cost is calculated in terms of Giga FLoating-point OPerations per second (GFLOPs). 

The results are shown in Table~\ref{tab:ablation}. The first row is our baseline, where no LASA operation is added. The second row shows the result of applying the LASA operation while disabling the gating mask. 
As can be seen, by adding the LASA operation, the PSNR has increased by 0.66dB, 0.30dB, 0.38dB on the scaling factors of 2, 3 and 4, respectively. 
Meanwhile, when performing the LASA operation, we observe a 66$\%$ increase in GFLOPs. 
However, comparing the results between the second row and the third row, we can find that, by utilizing the gating mask and limiting the computational resource, the GFLOPs decrease significantly while maintaining comparable and even better performance in PSNR and SSIM. 
The results demonstrate the effectiveness of our proposed LASA operation and gating mask. Meanwhile, the results verify the rationality that we should allocate different computational resources to areas of varying difficulty in SR tasks.

\begin{figure*}[t]
    \centering
    \includegraphics[width=0.99\textwidth]{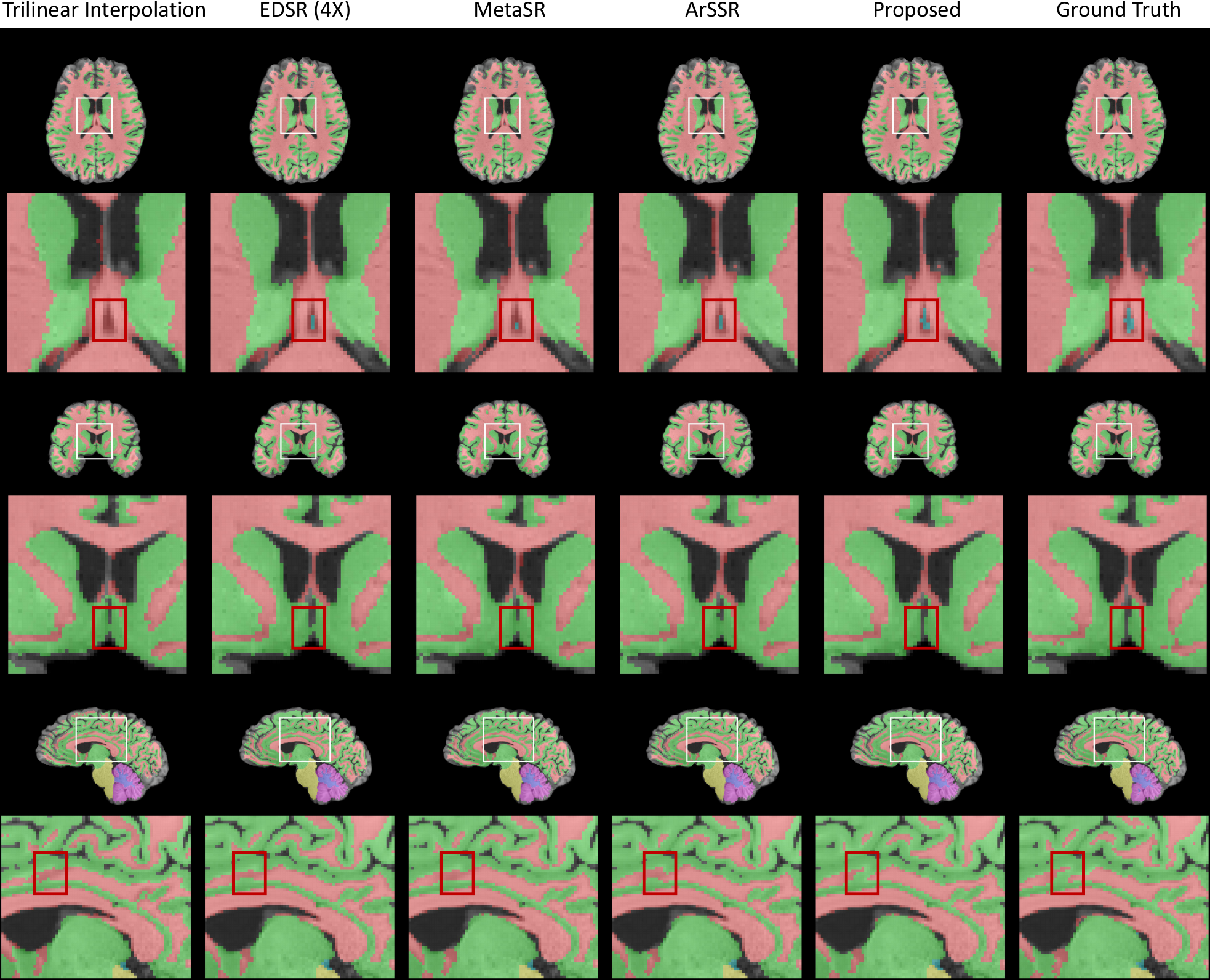}
    \caption{Visual comparison of the fully automatic segmentation on $\times4$ SR results by all the comparing models. The coronal, axial, and sagittal views are shown in three rows, respectively. The areas surrounded by the white boxes are zoomed in below. }
    \label{fig7}
\end{figure*}

\subsection{Experiments on Downstream Segmentation Task}
To further prove the effectiveness of our method, we follow~\citet{Chen2020MRISW} by evaluating our model on the brain segmentation task. 
For the segmentation toolkit, we use the Fastsurfer~\citep{fastsurfer}, an open-source and extensively used tool, to conduct fully automatic segmentation on the SR results from Section \ref{sec:sup_brain} and the ground-truth HR images. 
The Dice ratio is used to quantitatively evaluate the coherence in the segmentation results between the SR images and the ground-truth images. 
The higher the Dice ratio is, the segmentation results of SR images are closer to those of the HR images.

We calculate the Dice ratios for six tissues: Cortical White Matter, Cortical Grey Matter, Cerebellum White Matter, Cerebellum Cortex, Brain Stem, and Cerebrospinal Fluid (CSF). 
The quantitative comparisons are shown in Table~\ref{tab:seg}, we compare our method with trilinear interpolation, EDSR, MetaSR and ArSSR. In terms of the Dice ratio, the proposed SA-INR method constantly outperforms its counterparts on three different scaling factors (\textit{p}$<$0.01 with paired \textit{t}-tests). 

Also, we visualize the segmentation results of a test case on the scaling factor of 4.
We use pink, green, purple, blue, yellow and cyan to represent Cortical White Matter, Cortical Grey Matter, Cerebellum White Matter, Cerebellum Cortex, Brain Stem, and CSF, respectively. The results of three views (coronal, axial, and sagittal) are shown in Fig.~\ref{fig7}. As can be seen, the segmentation results of our model are closest to the ground truth.
From the first row, we could find that the regions of CSF have disappeared in the results of other methods, which are mostly preserved by our method.
From the second row, we find that the grey matter is mistakenly labeled in the images produced by other methods. Similarly, there is adhesion of the white matter in the third row, which may be caused by excessive smoothing in the SR images. 
In contrast, our method successfully preserves the initially separated structure.

\begin{figure*}[t]
    \centering
    \includegraphics[width=\textwidth]{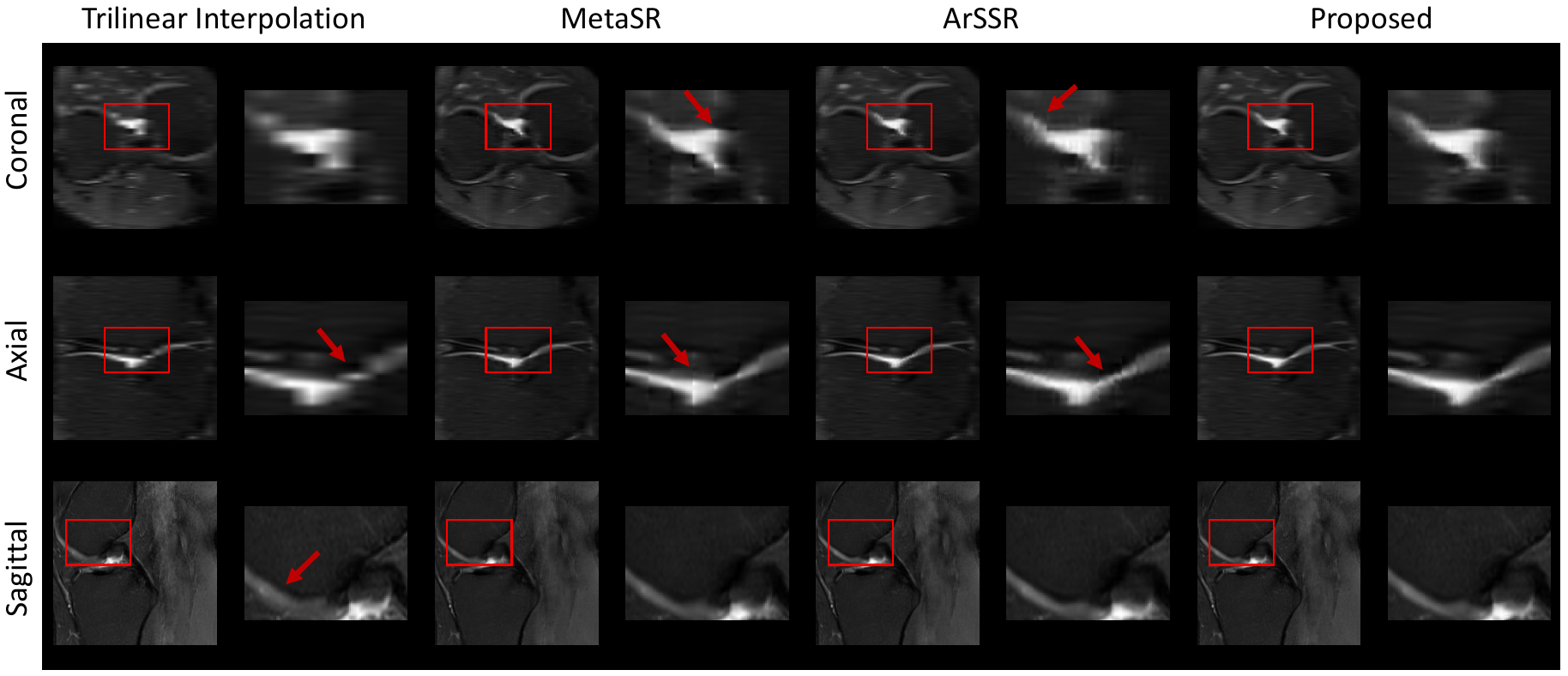}
    \caption{Visual comparisons of the isotropic knee SR results. The coronal, axial, and sagittal views are shown in three rows, respectively. The areas surrounded by the red boxes are zoomed in on the right side. The red arrows point out the artifacts.}
    \label{fig8}
\end{figure*}

\subsection{Experiments on Knee MRI (Without Ground Truth)}
\subsubsection{Dataset}
The previous simulation experiments show that our model performs better than other methods. 
Furthermore, it is crucial to show our model's capability in real-world clinical scenarios. 
Therefore, a total of 2,627 knee MRI subjects have been collected using fat-suppressed T2-weighted fast spin-echo (TSE) sequence. The intra-slice spacing is 0.3mm$\times$0.3mm, and the inter-slice spacing is 3.3mm (871 subjects) or 4.5mm (1756 subjects).

\subsubsection{Self-supervised Framework}
Since there is no ground truth, we follow~\citet{kai} by synthesizing HR-LR pairs first and then training our model. 
\begin{itemize}
    \item In stage I, we synthesize HR volumes given only LR volumes that are available. Specifically, we train a VAE~\citep{vae} using 2D in-plane slices from LR images.
    Then, we encode the slices of an LR volume and perform linear interpolation between their representations in the embedding space. 
    More slices can be acquired after processing these linearly interpolated representations through the decoder in VAE. 
    Finally, the HR images of reduced slice spacing are generated by stacking the decoded slices. 
    \item Although these synthesized HR images are realistic in appearance, they may still suffer from unexpected morphing induced by VAE, implying that the synthesized HR images cannot be paired with the real LR images. 
    However, we can degrade the synthesized HR images to generate the corresponding LR images and thus LR-HR pairs. Then we can train the supervised SR network based on these synthesized pairs.
\end{itemize}


\subsubsection{Visual Comparisons}

Unlike the brain dataset in the last experiment, images in the knee dataset lack HR ground truth. 
Therefore, we only provide visual comparisons of different SR methods Fig.~\ref{fig8}. Here we present the three views (coronal, axial, and sagittal) of a test case whose inter-slice spacing has been reduced from 4.5mm to 0.3mm ($\times$15).

In the first row of the figure, we display the coronal slices and zoom in on the anterior cruciate ligament (ACL). 
As can be seen, trilinear interpolation offers unacceptable image quality, making the ACL structure unrecognizable. The results of MetaSR and ArSSR are better but still have noticeable artifacts, as indicated by the red arrows. In comparison, the image generated by our proposed SA-INR method is smoother and clearer.
In the second row, we display the axial slices and zoom in on the synovial fluid and cartilage. The results of trilinear interpolation are incorrect, in which the cartilage is discontinuous. When coming to the MetaSR and ArSSR, the generated images are better but still contain artifacts. The proposed method performs better and provides a clear view of the tissues.
In the third row, we display the sagittal slices and zoom in on the articular (femur) cartilage. Trilinear interpolation generates blurry cartilage, making the diagnosis of diseases such as osteoarthritis difficult for both radiologists and computer-assisted diagnosis systems. Meanwhile, the rest of the three methods generate structurally plausible images. 

In summary, our proposed method shows distinct advantages over trilinear interpolation in the three views.
Besides, our method produces more realistic structures in coronal and axial views than MetaSR and ArSSR. The findings support our conclusion that the proposed SA-INR delivers better SR capability in real clinical cases.

\section{Conclusion and Discussion}

In this paper, we propose a novel deep-learning-based method, namely SA-INR for the arbitrary reduction of MR slice spacing. 
Instead of training a separate model for each scaling factor, SA-INR can handle different and non-integer scaling factors automatically. 
The key idea of SA-INR is to represent an LR image as a continuous implicit function of coordinates, such that the HR image with desired inter-slice spacing can be reconstructed based on the continuously upsampled coordinates in the 3D volume. 
Compared to other arbitrary-scale SR methods, SA-INR demonstrates it significant improvements in SR as well as the downstream segmentation task. 

Here we also present the limitations of our current work. 
First, SA-INR is only tested on the task of reducing MR slice spacing, while ignoring the issue of slice thickness in MR. Slice spacing reduction and slice thickness reduction are two different tasks, but they share the same basis of slice interpolation to a certain extent. Our work focuses only on slice spacing reduction, but does not consider the slice thickness. 
Future work will be conducted for more investigation to resolve this issue.
Second, when we evaluate the performance of SR approaches, we do not include an experiment that uses real pairs of LR/HR data. Obtaining such data is challenging in practice, which may need extremely precise registration between the LR and HR acquisitions.

\bibliographystyle{IEEEtran}
\bibliography{egbib.bib}

\begin{thebibliography}{10}
\providecommand{\url}[1]{#1}
\csname url@samestyle\endcsname
\providecommand{\newblock}{\relax}
\providecommand{\bibinfo}[2]{#2}
\providecommand{\BIBentrySTDinterwordspacing}{\spaceskip=0pt\relax}
\providecommand{\BIBentryALTinterwordstretchfactor}{4}
\providecommand{\BIBentryALTinterwordspacing}{\spaceskip=\fontdimen2\font plus
\BIBentryALTinterwordstretchfactor\fontdimen3\font minus
  \fontdimen4\font\relax}
\providecommand{\BIBforeignlanguage}[2]{{%
\expandafter\ifx\csname l@#1\endcsname\relax
\typeout{** WARNING: IEEEtran.bst: No hyphenation pattern has been}%
\typeout{** loaded for the language `#1'. Using the pattern for}%
\typeout{** the default language instead.}%
\else
\language=\csname l@#1\endcsname
\fi
#2}}
\providecommand{\BIBdecl}{\relax}
\BIBdecl

\bibitem{mri}
D.~Weishaupt, V.~D. K{\"o}chli, B.~Marincek, J.~M. Froehlich, D.~Nanz, and
  K.~P. Pruessmann, \emph{How does MRI work?: an introduction to the physics
  and function of magnetic resonance imaging}.\hskip 1em plus 0.5em minus
  0.4em\relax Springer, 2006, vol.~2.

\bibitem{ZHANG2017531}
J.~Zhang, L.~Zhang, L.~Xiang, Y.~Shao, G.~Wu, X.~Zhou, D.~Shen, and Q.~Wang,
  ``Brain atlas fusion from high-thickness diagnostic magnetic resonance images
  by learning-based super-resolution,'' \emph{Pattern Recognition}, vol.~63,
  pp. 531--541, 2017.

\bibitem{DeepResolve}
A.~S. Chaudhari, Z.~Fang, F.~Kogan, J.~Wood, K.~J. Stevens, E.~K. Gibbons,
  J.~H. Lee, G.~E. Gold, and B.~A. Hargreaves, ``Super-resolution
  musculoskeletal mri using deep learning,'' \emph{Magnetic resonance in
  medicine}, vol.~80, no.~5, pp. 2139--2154, 2018.

\bibitem{smore}
C.~Zhao, B.~E. Dewey, D.~L. Pham, P.~A. Calabresi, D.~S. Reich, and J.~L.
  Prince, ``Smore: a self-supervised anti-aliasing and super-resolution
  algorithm for mri using deep learning,'' \emph{IEEE transactions on medical
  imaging}, vol.~40, no.~3, pp. 805--817, 2020.

\bibitem{kai}
K.~Xuan, L.~Si, L.~Zhang, J.~Xue, Y.~Jiao, W.~Yao, D.~Wu, and Q.~Wang,
  ``Reducing magnetic resonance image spacing by learning without
  ground-truth,'' \emph{Pattern Recognition}, vol. 120, p. 108103, 06 2021.

\bibitem{TSCNet}
Z.~Lu, Z.~Li, J.~Wang, J.~Shi, and D.~Shen, ``Two-stage self-supervised
  cycle-consistency network for reconstruction of thin-slice mr images,'' in
  \emph{Medical Image Computing and Computer Assisted Intervention -- MICCAI
  2021}.\hskip 1em plus 0.5em minus 0.4em\relax Cham: Springer International
  Publishing, 2021, pp. 3--12.

\bibitem{nerf}
B.~Mildenhall, P.~P. Srinivasan, M.~Tancik, J.~T. Barron, R.~Ramamoorthi, and
  R.~Ng, ``Nerf: Representing scenes as neural radiance fields for view
  synthesis,'' in \emph{European conference on computer vision}.\hskip 1em plus
  0.5em minus 0.4em\relax Springer, 2020, pp. 405--421.

\bibitem{LIIF}
Y.~Chen, S.~Liu, and X.~Wang, ``Learning continuous image representation with
  local implicit image function,'' in \emph{2021 IEEE/CVF Conference on
  Computer Vision and Pattern Recognition (CVPR)}, 2021, pp. 8624--8634.

\bibitem{ArSSR}
Q.~Wu, Y.~Li, Y.~Sun, Y.~Zhou, H.~Wei, J.~Yu, and Y.~Zhang, ``An arbitrary
  scale super-resolution approach for 3-dimensional magnetic resonance image
  using implicit neural representation,'' 2021.

\bibitem{gumbel}
E.~Jang, S.~Gu, and B.~Poole, ``Categorical reparameterization with
  gumbel-softmax,'' \emph{arXiv preprint arXiv:1611.01144}, 2016.

\bibitem{vae}
D.~P. Kingma and M.~Welling, ``Auto-encoding variational bayes,'' \emph{arXiv
  preprint arXiv:1312.6114}, 2013.

\bibitem{SRCNN}
C.~Dong, C.~C. Loy, K.~He, and X.~Tang, ``Image super-resolution using deep
  convolutional networks,'' \emph{IEEE Transactions on Pattern Analysis and
  Machine Intelligence}, vol.~38, no.~2, pp. 295--307, 2016.

\bibitem{VDSR}
J.~Kim, J.~K. Lee, and K.~M. Lee, ``Accurate image super-resolution using very
  deep convolutional networks,'' in \emph{Proceedings of the IEEE conference on
  computer vision and pattern recognition}, 2016, pp. 1646--1654.

\bibitem{DRCN}
------, ``Deeply-recursive convolutional network for image super-resolution,''
  in \emph{Proceedings of the IEEE conference on computer vision and pattern
  recognition}, 2016, pp. 1637--1645.

\bibitem{EDSR}
B.~Lim, S.~Son, H.~Kim, S.~Nah, and K.~M. Lee, ``Enhanced deep residual
  networks for single image super-resolution,'' in \emph{2017 IEEE Conference
  on Computer Vision and Pattern Recognition Workshops (CVPRW)}, 2017, pp.
  1132--1140.

\bibitem{RDN}
Y.~Zhang, Y.~Tian, Y.~Kong, B.~Zhong, and Y.~Fu, ``Residual dense network for
  image super-resolution,'' in \emph{2018 IEEE/CVF Conference on Computer
  Vision and Pattern Recognition}, 2018, pp. 2472--2481.

\bibitem{RCAN}
Y.~Zhang, K.~Li, K.~Li, L.~Wang, B.~Zhong, and Y.~Fu, ``Image super-resolution
  using very deep residual channel attention networks,'' in \emph{Proceedings
  of the European Conference on Computer Vision (ECCV)}, September 2018.

\bibitem{chen2019learning}
Z.~Chen and H.~Zhang, ``Learning implicit fields for generative shape
  modeling,'' in \emph{Proceedings of the IEEE/CVF Conference on Computer
  Vision and Pattern Recognition}, 2019, pp. 5939--5948.

\bibitem{genova2020local}
K.~Genova, F.~Cole, A.~Sud, A.~Sarna, and T.~Funkhouser, ``Local deep implicit
  functions for 3d shape,'' in \emph{Proceedings of the IEEE/CVF Conference on
  Computer Vision and Pattern Recognition}, 2020, pp. 4857--4866.

\bibitem{derf}
D.~Rebain, W.~Jiang, S.~Yazdani, K.~Li, K.~M. Yi, and A.~Tagliasacchi, ``Derf:
  Decomposed radiance fields,'' in \emph{Proceedings of the IEEE/CVF Conference
  on Computer Vision and Pattern Recognition}, 2021, pp. 14\,153--14\,161.

\bibitem{occupancy}
L.~Mescheder, M.~Oechsle, M.~Niemeyer, S.~Nowozin, and A.~Geiger, ``Occupancy
  networks: Learning 3d reconstruction in function space,'' in
  \emph{Proceedings of the IEEE/CVF conference on computer vision and pattern
  recognition}, 2019, pp. 4460--4470.

\bibitem{convOccupancy}
S.~Peng, M.~Niemeyer, L.~Mescheder, M.~Pollefeys, and A.~Geiger,
  ``Convolutional occupancy networks,'' in \emph{European Conference on
  Computer Vision}.\hskip 1em plus 0.5em minus 0.4em\relax Springer, 2020, pp.
  523--540.

\bibitem{stanley2007compositional}
K.~O. Stanley, ``Compositional pattern producing networks: A novel abstraction
  of development,'' \emph{Genetic programming and evolvable machines}, vol.~8,
  no.~2, pp. 131--162, 2007.

\bibitem{sitzmann}
V.~Sitzmann, J.~Martel, A.~Bergman, D.~Lindell, and G.~Wetzstein, ``Implicit
  neural representations with periodic activation functions,'' \emph{Advances
  in Neural Information Processing Systems}, vol.~33, pp. 7462--7473, 2020.

\bibitem{bemana2020x}
M.~Bemana, K.~Myszkowski, H.-P. Seidel, and T.~Ritschel, ``X-fields: Implicit
  neural view-, light-and time-image interpolation,'' \emph{ACM Transactions on
  Graphics (TOG)}, vol.~39, no.~6, pp. 1--15, 2020.

\bibitem{Karras2021}
T.~Karras, M.~Aittala, S.~Laine, E.~H\"ark\"onen, J.~Hellsten, J.~Lehtinen, and
  T.~Aila, ``Alias-free generative adversarial networks,'' in \emph{Proc.
  NeurIPS}, 2021.

\bibitem{MetaSR}
X.~Hu, H.~Mu, X.~Zhang, Z.~Wang, T.~Tan, and J.~Sun, ``Meta-sr: A
  magnification-arbitrary network for super-resolution,'' in \emph{2019
  IEEE/CVF Conference on Computer Vision and Pattern Recognition (CVPR)}, 2019,
  pp. 1575--1584.

\bibitem{IREM}
Q.~Wu, Y.~Li, L.~Xu, R.~Feng, H.~Wei, Q.~Yang, B.~Yu, X.~Liu, J.~Yu, and
  Y.~Zhang, ``Irem: High-resolution magnetic resonance image reconstruction via
  implicit neural representation,'' in \emph{Medical Image Computing and
  Computer Assisted Intervention -- MICCAI 2021}, 2021, pp. 65--74.

\bibitem{NL}
X.~Wang, R.~Girshick, A.~Gupta, and K.~He, ``Non-local neural networks,'' in
  \emph{2018 IEEE/CVF Conference on Computer Vision and Pattern Recognition},
  2018, pp. 7794--7803.

\bibitem{chang2019differentiable}
J.~Chang, X.~Zhang, Y.~Guo, G.~Meng, S.~Xiang, and C.~Pan, ``Differentiable
  architecture search with ensemble gumbel-softmax,'' \emph{arXiv preprint
  arXiv:1905.01786}, 2019.

\bibitem{herrmann2020channel}
C.~Herrmann, R.~S. Bowen, and R.~Zabih, ``Channel selection using gumbel
  softmax,'' in \emph{European Conference on Computer Vision}.\hskip 1em plus
  0.5em minus 0.4em\relax Springer, 2020, pp. 241--257.

\bibitem{Verelst_2020}
T.~Verelst and T.~Tuytelaars, ``Dynamic convolutions: Exploiting spatial
  sparsity for faster inference,'' in \emph{2020 {IEEE}/{CVF} Conference on
  Computer Vision and Pattern Recognition ({CVPR})}.\hskip 1em plus 0.5em minus
  0.4em\relax {IEEE}, jun 2020.

\bibitem{HCP}
M.~F. Glasser, S.~N. Sotiropoulos, J.~A. Wilson, T.~S. Coalson, B.~Fischl,
  J.~L. Andersson, J.~Xu, S.~Jbabdi, M.~Webster, J.~R. Polimeni, D.~C. {Van
  Essen}, and M.~Jenkinson, ``The minimal preprocessing pipelines for the human
  connectome project,'' \emph{NeuroImage}, vol.~80, pp. 105--124, 2013, mapping
  the Connectome.

\bibitem{fastsurfer}
L.~Henschel, S.~Conjeti, S.~Estrada, K.~Diers, B.~Fischl, and M.~Reuter,
  ``Fastsurfer-a fast and accurate deep learning based neuroimaging pipeline,''
  \emph{NeuroImage}, vol. 219, p. 117012, 2020.

\bibitem{PSNR}
Q.~Huynh-Thu and M.~Ghanbari, ``Scope of validity of psnr in image/video
  quality assessment,'' \emph{Electronics letters}, vol.~44, no.~13, pp.
  800--801, 2008.

\bibitem{SSIM}
A.~Hore and D.~Ziou, ``Image quality metrics: Psnr vs. ssim,'' in \emph{2010
  20th international conference on pattern recognition}.\hskip 1em plus 0.5em
  minus 0.4em\relax IEEE, 2010, pp. 2366--2369.

\bibitem{s3}
C.~T. Vu and D.~M. Chandler, ``S3: a spectral and spatial sharpness measure,''
  in \emph{2009 First International Conference on Advances in
  Multimedia}.\hskip 1em plus 0.5em minus 0.4em\relax IEEE, 2009, pp. 37--43.

\bibitem{pytorch}
A.~Paszke, S.~Gross, S.~Chintala, G.~Chanan, E.~Yang, Z.~DeVito, Z.~Lin,
  A.~Desmaison, L.~Antiga, and A.~Lerer, ``Automatic differentiation in
  pytorch,'' in \emph{NIPS 2017 Workshop on Autodiff}, 2017.

\bibitem{Adam}
D.~P. Kingma and J.~Ba, ``Adam: A method for stochastic optimization,'' 2014,
  cite arxiv:1412.6980Comment: Published as a conference paper at the 3rd
  International Conference for Learning Representations, San Diego, 2015.

\bibitem{Chen2020MRISW}
Y.~Chen, A.~G. Christodoulou, Z.~Zhou, F.~Shi, Y.~Xie, and D.~Li, ``Mri
  super-resolution with gan and 3d multi-level densenet: Smaller, faster, and
  better,'' \emph{ArXiv}, vol. abs/2003.01217, 2020.

\end{thebibliography}

\end{document}